\documentclass[preprint,12pt]{elsarticle}

\usepackage[protrusion=true,expansion=false]{microtype} %% Fixes overfull hboxes / margin overflows
\usepackage{amssymb}
\usepackage{amsmath}
\usepackage{siunitx}
\usepackage{hyperref}
\usepackage{url}
\usepackage{booktabs}
\usepackage[table]{xcolor}
\biboptions{sort&compress}
\usepackage{lmodern}

\journal{Optics \& Laser Technology}

\begin{document}
	
	\begin{frontmatter}
		
		%\title{Azimuthal super-pupil beam engineering for improved fluorescence depletion microscopy}
		\title{\textcolor{black}{Super-pupil beam engineering for enhanced confinement of azimuthally polarized depletion beams}} %%% Jordi   R1.1

		%% Authors and affiliations using elsarticle commands
		\author[ub,in2ub]{Costanza Agazzi\fnref{equal}}
		\author[ub,in2ub]{Nick Toledo-Garc\'ia\fnref{equal}}
		\author[ub,in2ub]{Estela Mart\'in-Badosa}
		\author[ub,in2ub]{Mario Montes-Usategui}
		\author[ub,in2ub]{David Maluenda}
		\author[ub,in2ub]{Jordi Tiana-Alsina}
		\author[ucm]{Rosario Mart\'inez-Herrero}
		\author[ub]{Artur Carnicer\corref{cor1}}
		\ead{artur.carnicer@ub.edu}

		\cortext[cor1]{Corresponding author}
		\fntext[equal]{These authors contributed equally to this work.}
		
		\affiliation[ub]{organization={Departament de F\'isica Aplicada, Facultat de F\'isica, Universitat de Barcelona},
			addressline={Carrer de Mart\'i i Franqu\`es 1},
			city={Barcelona},
			postcode={08028},
			country={Spain}}
		
		\affiliation[in2ub]{organization={Institut de Nanoci\`encia i Nanotecnologia (IN2UB), Universitat de Barcelona},
			postcode={08028},
			country={Spain}}
		
		\affiliation[ucm]{organization={Departamento de \'Optica, Facultad de Ciencias F\'isicas, Universidad Complutense de Madrid},
			addressline={Plaza de las Ciencias 1},
			city={Madrid},
			postcode={28040},
			country={Spain}}
		
		%% Abstract
		\begin{abstract}
			Fluorescence depletion microscopy techniques
			%such as STED and RESOLFT    %    R1.1
 			require optical fields with a well-defined and spatially confined central intensity minimum to achieve sub-diffraction lateral resolution. Here, we present the design and experimental implementation of an azimuthally polarized, doughnut-shaped depletion beam based on super-pupil engineering principles. By tailoring the radial amplitude distribution at the entrance pupil to approximate a Bessel-type target function, the resulting focal field exhibits a tighter central doughnut compared to conventional circularly or azimuthally polarized beams. The designed pupil field distribution is implemented using a phase-only spatial light modulator operated in a double-pass configuration, enabling independent modulation of orthogonal polarization components via complex-field holographic encoding.
            Analytical modeling of the designed field predicts an estimated resolution gain of approximately \textcolor{black}{24\%} and a reduction in required depletion power of approximately \textcolor{black}{42\%} relative to a canonical circularly polarized Laguerre-Gaussian beam.
			Although the engineered field exhibits pronounced sidelobes, these do not preclude its use as a depletion beam, since lateral resolution is strongly influenced by the spatial confinement and effective suppression of the central intensity minimum for a given depletion intensity.
			%This suggests that the proposed approach can enable improved lateral resolution at comparable depletion powers, providing a flexible and experimentally accessible route for engineering depletion fields in reconfigurable super-resolution microscopy systems.
			\textcolor{black}{This suggests that the proposed beam may provide
			improved depletion efficiency through enhanced focal confinement,
            offering a flexible and experimentally accessible approach to depletion
			beam engineering for reconfigurable superresolution microscopy systems.} %%% Jordi     R1.1
		\end{abstract}
		
		%%Graphical abstract
		%\begin{graphicalabstract}
			%\includegraphics{grabs}
		%\end{graphicalabstract}
		
		%%Research highlights
		\begin{highlights}
			\item Super-pupil engineering enhances confinement of azimuthally polarized doughnut beams.  %%% TODO: Almost same than title!!! ************************
			\item Analytical model predicts 24\% resolution gain and 42\% power reduction vs Laguerre-Gaussian beams.
			\item Double-pass SLM enables independent vectorial polarization control.
			\item Pixel pooling mitigates SLM cross-talk and deepens on-axis null.
			\item Sub-diffraction bead scans confirm predicted focal beam reshaping.
		\end{highlights}
		
		%% Keywords
		\begin{keyword}
			Vector beams \sep azimuthal polarization \sep pupil engineering \sep spatial light modulator \sep super-resolution \sep STED microscopy
		\end{keyword}
		
	\end{frontmatter}
	
	%% Uncomment to enable line numbers:
	%% \linenumbers
	
	%% ===================================================================
	\section{Introduction}
	\label{sec:intro}
	%% ===================================================================
	
	For over a century, the diffraction limit defined the spatial resolution boundary in optical microscopy. The emergence of super-resolution methods has fundamentally changed this paradigm by introducing techniques that modulate excitation patterns to circumvent classical optical constraints.
	Among these, depletion-based techniques such as stimulated emission depletion (STED) and reversible saturable optical fluorescence transitions (RE\-SOLFT) achieve resolution enhancement through the controlled suppression of fluorescence.
	
	In these techniques, fluorescence is selectively inhibited in the outer regions of the excitation focus using a light field with a well-defined central intensity minimum \cite{Sahl2019l, Hell:1995aa, Hofmann:2005aa}. The efficiency and spatial confinement of this depletion process critically depend on the beam's intensity and distribution at the focus. More generally, the achievable resolution in STED microscopy is governed by both the applied depletion intensity and the spatial characteristics of the central intensity minimum, so that improving the confinement of the zero can reduce the intensity required to reach a given resolution \cite{Galiani:12}.
	 In most STED implementations, circularly polarized beams are employed because they are easy to generate and provide an effective doughnut-shaped intensity profile when combined with a vortex phase mask \cite{Harke:08,Nepaune:2013}. Nevertheless, the depletion efficiency can be further optimized by employing vector beams with cylindrical symmetry, such as radially or azimuthally polarized modes. %, which enable tighter focusing and improved polarization control at high numerical apertures~\cite{dorn2003sharper,zhan2009cylindrical,Youngworth:00, Yang:2016, Hao:2010}. %Beyond their favorable vectorial focusing properties,
	\textcolor{black}{Azimuthally polarized beams have attracted considerable interest as depletion beams because they naturally generate a doughnut-shaped focal distribution with a more tightly confined central intensity minimum than conventional circularly polarized Laguerre–Gaussian (LG) beams under high-numerical-aperture focusing. Furthermore, the central dark region of azimuthally polarized beams has been reported to remain substantially better preserved in the presence of primary astigmatism than that of conventional LG beams, making them attractive candidates for advanced depletion beam engineering~\cite{Hao:2010, singh2009tight}. Conversely, the strong sensitivity of conventional LG depletion beams to low-order optical aberrations, particularly astigmatism, has been extensively documented~\cite{deng2010effects, ohland2019study}.
	The practical relevance of aberration-induced degradation in LG depletion beams has also been evidenced experimentally in parallel subtraction microscopy, where the quality of the generated doughnut beams progressively deteriorates towards the edge of the field of view because of off-axis aberrations~\cite{klingmann2024parallel}.}  %%% Jordi  R2.8
	
	The generation of such beams with arbitrary polarization and amplitude distributions has been extensively studied using interferometric and holographic approaches based on spatial light modulators (SLMs). These systems enable independent modulation of orthogonal polarization components and the synthesis of focused fields with prescribed vectorial properties~\cite{Moreno:12,Ostrovsky:13,maluenda2013reconfigurable,maluenda2014synthesis,herrero18needle}. Experimental realizations have demonstrated the generation of radially, azimuthally, and circularly polarized focal fields, as well as advanced shaping of the focal volume by tailoring the phase and amplitude in the pupil plane~\cite{maluenda2014synthesis,herrero18needle}.
	
	The concept of super-pupils (i.e.\ pupil filters designed to engineer the angular spectrum of the illumination) extends this idea by enabling the formation of sub-diffraction or extended-depth focal distributions.
    By properly controlling the amplitude and phase across the pupil, super-pupil designs can produce superoscillatory or Bessel-type beams characterized by narrow central minima and elongated propagation ranges~\cite{kitamura2010sub,wang2008creation,Chen2019}.
    \textcolor{black}{The extended propagation range makes these beams particularly attractive for preserving the beam profile along the optical axis, thereby maintaining their super-resolution performance in thick samples~\cite{yu2016super}.}  %%% Artur   R2.1
    These beams have been extensively investigated for applications in high-resolution microscopy, where they can produce optical needles or extended focal regions~\cite{herrero18needle,kitamura2010sub}.
	
	However, their direct application to imaging techniques, such as confocal or structured illumination microscopy (SIM), has remained limited because the strong sidelobes inherent to superoscillatory fields degrade the point-spread function and compromise image reconstruction~\cite{Chen2019,Wong2013}. In contrast, these same beams can be particularly advantageous for fluorescence depletion microscopy, where the achievable lateral resolution is primarily influenced by the properties of the central zero-intensity region, while sidelobes have a comparatively minor impact on the detected signal. It should be noted, however, that while sidelobes do not degrade spatial resolution in this context, they do redistribute optical energy away from the depletion ring, which can reduce the total power available for fluorescence suppression. This redistribution introduces a trade-off between central confinement and depletion efficiency that must be considered when evaluating the practical performance of the engineered beam.
	\textcolor{black}{Rather than engineering a depletion beam from a conventional Laguerre–Gaussian mode, the present work builds upon the favorable focusing properties of azimuthally polarized beams and further enhances the confinement of their central intensity minimum through super-pupil engineering.}  %%% Jordi   R1.1   (centrar en beam-engineering)

	In this contribution we present an engineered vector beam, which advances conventional doughnut-shaped and vector depletion beams by combining pupil-engineered superoscillatory design with full vectorial polarization control in a reconfigurable, double-pass SLM-based platform. Unlike standard vortex or annular beams, whose central intensity minimum is primarily determined by the imposed topological charge and is highly sensitive to aberrations, our proposal optimizes the spatial confinement through a tailored pupil modulation \cite{deng2010effects, ohland2019study}. This strategy yields an improved on-axis intensity suppression consistent with a more confined central minimum, as experimentally confirmed by focal scans using sub-diffraction fluorescent beads, where a significantly sharper and deeper zero is revealed. Importantly, the work demonstrates that practical limitations of phase-only SLMs (such as pixel cross-talk and finite modulation fidelity) can be mitigated through controlled pixel pooling, enabling reliable reproduction of the designed pupil function while preserving the essential superoscillatory features of the beam. From an application perspective, the beam provides a tunable, experimentally accessible depletion field that can be integrated into existing STED or RESOLFT architectures, offering improved depletion contrast, enhanced robustness against implementation imperfections or aberrations, and the flexibility required for adaptive or parallelized super-resolution microscopy.
	 In this context, the present work focuses specifically on improving the lateral confinement of the depletion beam. The axial characteristics of the field, although relevant for three-dimensional imaging performance, are not addressed here.
	
	The paper is organized as follows: Section~\ref{sec:design} describes the mathematical concepts necessary for designing azimuthally polarized, highly focused, doughnut-shaped beams. Section~\ref{sec:params} discusses how to select the numerical parameters required to produce successful and implementable super-resolution depletion beams. Section~\ref{sec:improvements} presents estimated parameter improvements. Section~\ref{sec:experiment} presents the optical setup and provides an experimental verification of the generated beams. Results are discussed in Section~\ref{sec:discussion} and finally, Section~\ref{sec:conclusion} summarizes the conclusions.
	
	%% ===================================================================
	\section{Design of azimuthally polarized, highly focused doughnut-shaped beams}
	\label{sec:design}
	%% ===================================================================
	
\begin{figure}[h!]
		\centering
		\includegraphics[width=0.8\linewidth]{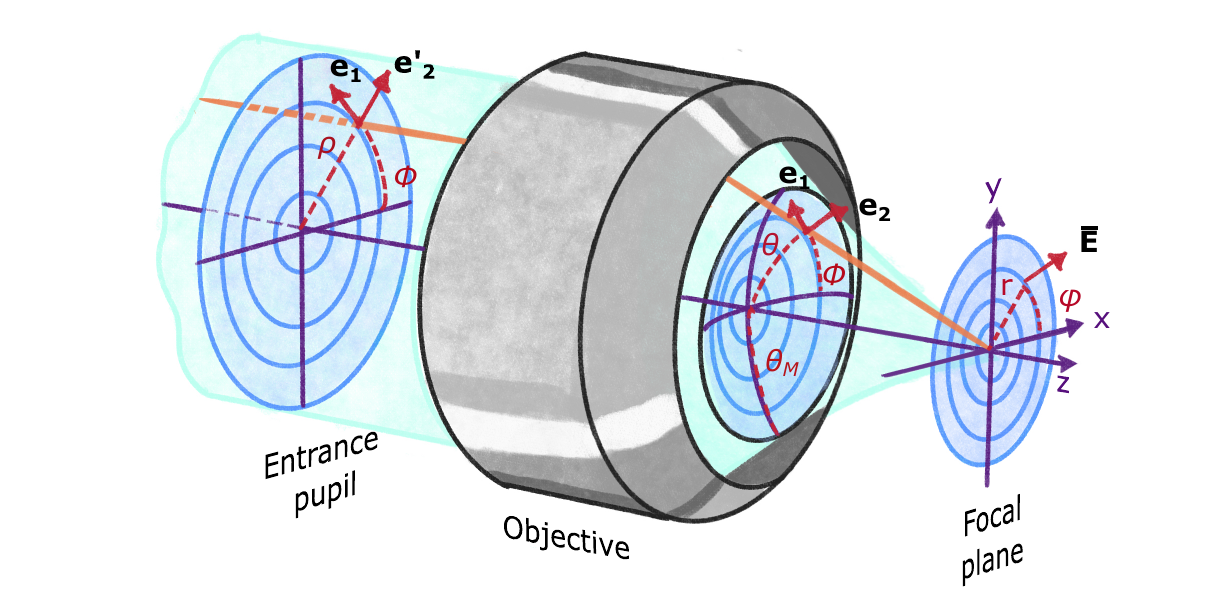}
		\caption{Reference coordinate systems and the variables involved in the process (adapted from \cite{martinez2020uncertainty}).}
		\label{fig:coordinates}
	\end{figure}

	The Richards-Wolf formula \cite{richards1959electromagnetic} describes the behavior of a focused electromagnetic beam in the focal region of a high numerical aperture (NA) lens. This equation relates the transverse illuminating beam $\mathbf{E_0} =(E_{0x},E_{0y},0)$ and the focused field at a distance $z$ from the focal plane $\mathbf{E}(r,\phi,z)$:
	\begin{align}
		\label{eq:RW1}
		\mathbf{E}(r,\varphi,z) \propto \int_0^{\theta_{\mathrm{M}} } \int_0^{2\pi } &\mathbf{E_\infty}  (\theta ,\phi )
		\exp \left (ikr\sin\theta \cos (\phi  - \varphi ) \right ) \cdot \nonumber \\
		&\exp \left( -ikz \cos\theta \right)\sin\theta \,d\theta \,d\phi\, ,
	\end{align}
	where $\mathbf{r}=(r,\varphi,z)$ are the coordinates in the focal area, $k$ is the wave number in a medium with refractive index $n$ ($k = 2\pi n / \lambda$), $\theta_M$ is the semi-aperture angle with $a = \sin\theta_M$, and $\theta$ and $\phi$ are the coordinates at the Gaussian sphere of reference. $\mathbf{E_\infty}$ is the so-called vector angular spectrum:
	\begin{equation}
		\mathbf{E_\infty} = P(\theta)\left(\left(\mathbf{E_0} \cdot \mathbf{e}_1\right)\mathbf{e}_1\, + \, \left(\mathbf{E_0} \cdot \mathbf{e}_2^i\right) \mathbf{e}_2\right),
		\label{eq:16}
	\end{equation}
	$f_1 = \mathbf{E_0} \cdot \mathbf{e}_1$ and $f_2 = \mathbf{E_0} \cdot \mathbf{e}_2^i$ are the azimuthal and radial transverse components of the incident transverse field $\mathbf{E_0}$, respectively. Vectors $\mathbf{e}_1$, $\mathbf{e}_2$ and $\mathbf{e}_2^{i}$ are described by:
	\begin{subequations}
		\begin{align}
			\mathbf{e}_1 (\phi ) &= ( - \sin \phi ,\cos \phi ,0) \\
			\mathbf{e}_2^{i} (\phi ) &= (\cos \phi ,\sin \phi ,0) \\
			\mathbf{e}_2 (\phi ,\theta ) &= ( \cos\theta \cos \phi ,\cos\theta \sin \phi ,\sin\theta).
		\end{align}
	\end{subequations}
	$P(\theta)$ is the so-called apodization function that takes the form $P(\theta)=\sqrt{\cos\theta}$ for isoplanatic systems. Figure~\ref{fig:coordinates} summarizes the systems of coordinates used in the entrance pupil, the Gaussian sphere of reference, and the focal plane.
	
	Let us consider the special case of an azimuthally polarized input beam with circular symmetry and no topological charge. For convenience, we perform a change of variables $\rho = \sin\theta$. Under these conditions, the input beam is expressed as \textcolor{black}{$\mathbf{E_0}(\rho) = \rho\, g(\rho)\, (-\sin\phi,\cos\phi,0)$}, %%% Artur   R1.3
    where $g(\rho)$ describes the amplitude profile. Note that $g(\rho)$ is multiplied by $\rho$ to guarantee $\mathbf{E_0}(0)=0$, thus avoiding an indeterminate polarization state at $\rho = 0$. The apodization function becomes $P(\rho) = (1-\rho^2)^{1/4}$. After some algebra, Eq.~(\ref{eq:RW1}) reads:
	\begin{equation}
		\mathbf{E}(r,\varphi,z) \propto \begin{pmatrix} -\sin\varphi \\ \cos\varphi \\ 0 \end{pmatrix}   \int_0^{a} \rho g(\rho) \frac{1}{(1-\rho^2)^{1/4}} \mathrm{J}_1(kr\rho) \exp \left ( -ikz \sqrt{1-\rho^2}\right )\rho d\rho .
		\label{eq:5}
	\end{equation}
	Defining the function $\epsilon (\rho)$ as $\epsilon(\rho) = \rho g(\rho) \dfrac{1}{(1-\rho^2)^{1/4}}$ and using the change of variables $t = \rho / a$, Eq.~(\ref{eq:5}) at $z=0$ becomes:
	\begin{equation}
		\mathbf{E}(r,\varphi,0) \propto \begin{pmatrix} -\sin\varphi \\ \cos\varphi \\ 0 \end{pmatrix}   \int_0^{1} \epsilon(at) \mathrm{J}_1(krat) t dt.
		\label{eq:axJ1}
	\end{equation}
	Because the function $\epsilon(at)$ can be expanded in terms of the radial Zernike polynomials $R_n^1(t)$
	\begin{equation}
		\epsilon(at) = \sum_{n \;\mathrm{odd}} \alpha_n R_n^1(t),
		\label{eq:epsilon}
	\end{equation}
	we can take advantage of the analytical relationship that links radial Zernike polynomials and Bessel functions \cite{richards1959electromagnetic, dai2006zernike}:
	\begin{equation}
		\int_0^1 R_n^1(p) J_1(pt) p dp = (-1)^{\frac{n-1}{2}}\;  \frac{\mathrm{J}_{n+1}(t)}{t}.   %%% Artur   R1.4
	\end{equation}
	Accordingly, the irradiance in the focal plane becomes:
	\begin{equation}
		I(r,0) = \left | a^2 \sum_{n \;\mathrm{odd}} \alpha_n \; (-1)^{\frac{n-1}{2}} \;\frac{\mathrm{J}_{n+1}(k a r)}{kar}\right |^2.
		\label{eq:intensitat}
	\end{equation}
	
	In this paper, we focus on designing beams with a profile $|\mathrm{J}_1(s n r)|^2$ in the focal plane, where $s$ is a scaling factor that allows us to adjust the beam width. The choice of $|\mathrm{J}_1(s n r)|^2$ as the target function is justified by its oscillatory character, which helps produce hollow beams with improved resolution \cite{BerryPopescu2006, Huang2007}.
	
	In order to obtain this target profile,
	the coefficients $\alpha_n$ in Eq.~\ref{eq:intensitat} are determined using the SciPy \cite{2020SciPy-NMeth} implementation of the differential evolution (DE) optimization method \cite{storn1997differential}. DE is well suited for global optimization tasks, helping to avoid local minima. In this problem, DE searches for the set of values $\alpha_n$ that minimize the distance between $I(r, 0)$ and a predefined target function. Once the set $\{\alpha_n\}$ is determined, the distribution $\rho g(\rho)$ at the entrance pupil of the microscope lens is obtained by
	\begin{equation}
		\rho g(\rho) = (1-\rho^2)^{1/4}\,\epsilon(\rho) = (1-\rho^2)^{1/4}\,\sum_{n\;\mathrm{odd}} \alpha_n R_n^1(\rho/a).
		\label{eq:rhoxg}
	\end{equation}
	
\begin{figure}[hb!]
		\centering
		\includegraphics[width=0.9\linewidth]{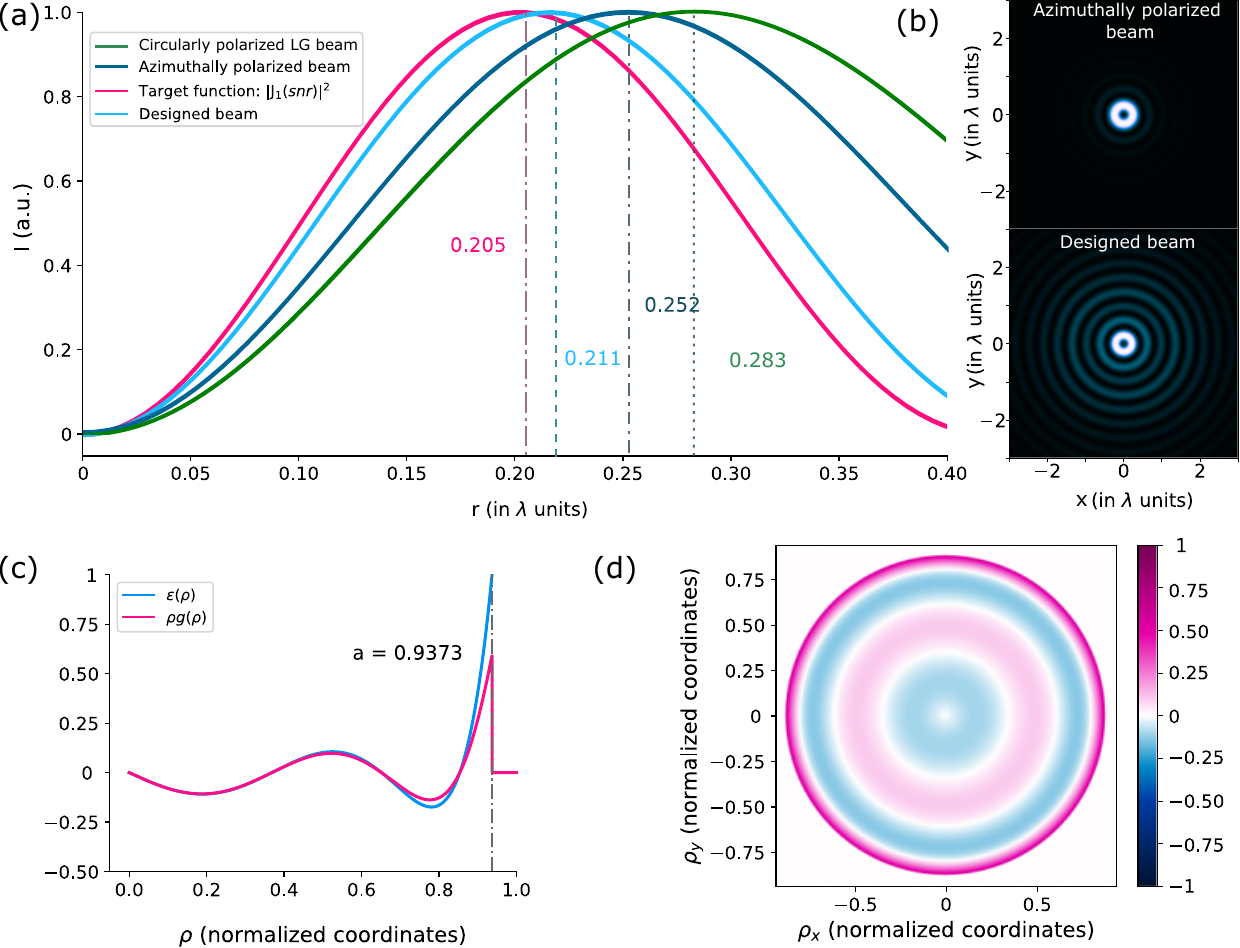}
		\caption{(a) Profiles of the target beam $|\mathrm{J}_1(s n r)|^2$ with $s=6$ (magenta curve), the designed beam (cyan curve), the azimuthally polarized beam (dark blue curve), and, for reference, a circularly polarized Laguerre-Gaussian beam (green curve). The vertical dotted lines and the labels inside the figure indicate the positions of the respective maxima. (b) 2D irradiance distributions of the top-hat azimuthally polarized beam and the designed beam in the focal plane. (c) Representation of the functions $\epsilon(\rho)$ and $\rho\,g(\rho)$. The vertical line indicates the value of $a$ (= NA$/n$). (d) Cartesian representation of the corresponding $\rho\,g(\rho)$ hologram.}
		\label{fig:azvsown}
	\end{figure}

	Figure~\ref{fig:azvsown}(a) compares the focal intensity profile of the target function $|\mathrm{J}_1(s n r)|^2$ with a scaling factor $s = 6$, against that of a top-hat azimuthally polarized beam and, for reference, a conventional circularly polarized Laguerre-Gaussian beam.
	\textcolor{black}{Throughout this work, the term top-hat beam refers to a beam with an approximately uniform amplitude distribution at the entrance pupil after holographic beam shaping. Although the laser illuminating the SLM has a Gaussian profile, the holographic encoding compensates for this apodization to produce the desired pupil-field distribution.}  %%% Jordi    R1.5
	
	This $s$ value represents a trade-off between resolution improvement and implementation complexity, as it will be discussed in Section~\ref{sec:params}. The reference beam is generated considering an oil-immersion objective lens with NA$=1.44$ ($n=1.515$, $a = 0.9373$). We use this as our benchmark because it produces a narrower doughnut-shaped beam than standard circular polarization, which is the typical depletion beam used in STED microscopy \cite{Sahl2019l, Hell:1995aa}.
	In the figure, the target function (magenta curve) and the reference beam (dark blue curve) reach their maxima at $r = 0.205\lambda$ and $r = 0.235\lambda$, respectively. The cyan curve, labelled \textit{Designed beam}, corresponds to a beam calculated using Eq.~(\ref{eq:intensitat}) and coefficients $\alpha_1=0.55, \alpha_3=0.92, \alpha_5=1,$ and $\alpha_7=0.67$. The rationale for selecting these specific values is detailed in the next section. For completeness, Fig.~\ref{fig:azvsown}(b) illustrates the irradiance distribution of both the designed beam and the top-hat azimuthally polarized beam in the focal plane. Finally, Fig.~\ref{fig:azvsown}(c) presents the designed functions $\epsilon(\rho)$ and $\rho g(\rho)$, along with the corresponding hologram (Fig.~\ref{fig:azvsown}(d)), obtained from Eq.~(\ref{eq:rhoxg}).

	%% ===================================================================
	\section{Estimation of the numerical parameters for designing the beam}
	\label{sec:params}
	%% ===================================================================
	
\begin{figure}[th!]
		\centering \includegraphics[width=0.8\linewidth]{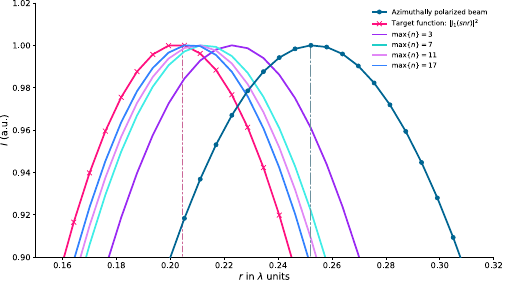}
		\caption{Profiles of $I(r,0)$ for max(n) = \{1, 7, 11, and 17\}, the azimuthally polarized {top-hat} beam (dark blue) and the target function (magenta). As in \hyperref[fig:azvsown]{Fig.~\ref{fig:azvsown}}, the vertical discontinuous lines indicate the position of the maxima of the azimuthally polarized beam and the target function.}
		\label{fig:Zer-z}
	\end{figure}

	The design of the proposed beam strongly depends on two related parameters: (i) the number of terms $n$ used in the calculation of $I(r,0)$, and (ii) the length of the design window $L$ (the domain of $r$ where the optimization takes place). Clearly, the larger the number of terms used in Eq.~\ref{eq:epsilon}, the better the approximation. The series converges rapidly, as shown in Fig.~\ref{fig:Zer-z}.
	
\begin{figure}[h!]
		\centering\includegraphics[width=0.8\linewidth]{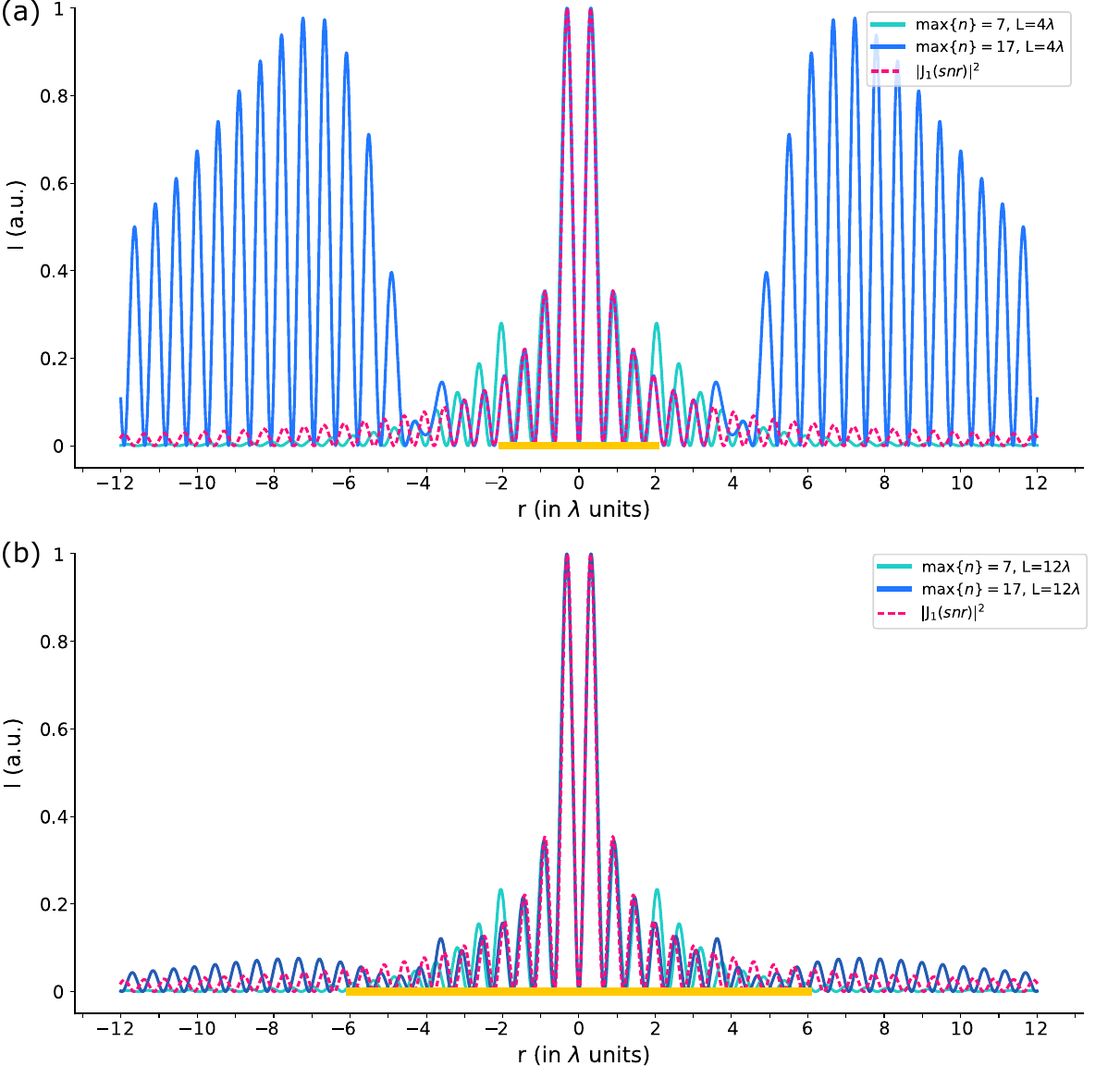}
		\caption{Dependence of the beam profile \textcolor{black}{on} the number of terms and the window size. (a) Window size $L = 4 \lambda$; (b) window size $L = 12 \lambda$. In both subfigures, $I(r,0)$ for $\mathrm{max}(n) = 7$ and $\mathrm{max}(n) = 17$ are depicted in cyan and blue, respectively. The horizontal, yellow line indicates the domain $L$ where the optimization takes place.}   %%% David    R1.8
		\label{fig:nterms_ws}
	\end{figure}

	Figure~\ref{fig:nterms_ws} illustrates the effect of $L$ and $\mathrm{max}(n)$ on the calculation of the coefficients $\{\alpha_n\}$. In Fig.~\ref{fig:nterms_ws}(a), the set $\{\alpha_n\}$ is obtained for $r \in [-2\lambda, 2\lambda]$ ($L=4\lambda$), whereas in Fig.~\ref{fig:nterms_ws}(b), $r \in [-6\lambda, 6\lambda]$ ($L=12\lambda$). Both subfigures display three curves: the calculated profiles $I(r,0)$ at the focal plane using 7 and 17 terms (cyan and blue curves, respectively), and the target function $|\mathrm{J}_1(s n r)|^2$ (magenta curve). Note that the cyan and blue curves are calculated using either Eqs.~(\ref{eq:5}) or (\ref{eq:intensitat}) and are displayed over the range $r \in [-12\lambda, 12\lambda]$. For the $L = 4\lambda$ case, $I(r,0)$ matches the target function $|\mathrm{J}_1(s n r)|^2$ within $[-2\lambda, 2\lambda]$, but for $|r| > 2\lambda$, the $I(r,0)$ curves exhibit sidelobes that are more pronounced for the $\mathrm{max}(n) = 17$ case. If the optimization is performed in a larger domain ($L=12$), as shown in Fig.~\ref{fig:nterms_ws}(b), sidelobes are less significant and the behavior of the $\mathrm{max}(n) = 7$ case is slightly better. Note that the appearance of sidelobes is a well-known effect when oscillating beams are focused \cite{Hyvarinen:12}.
	
\begin{figure}[ht!]
		\centering \includegraphics[width=0.95\linewidth]{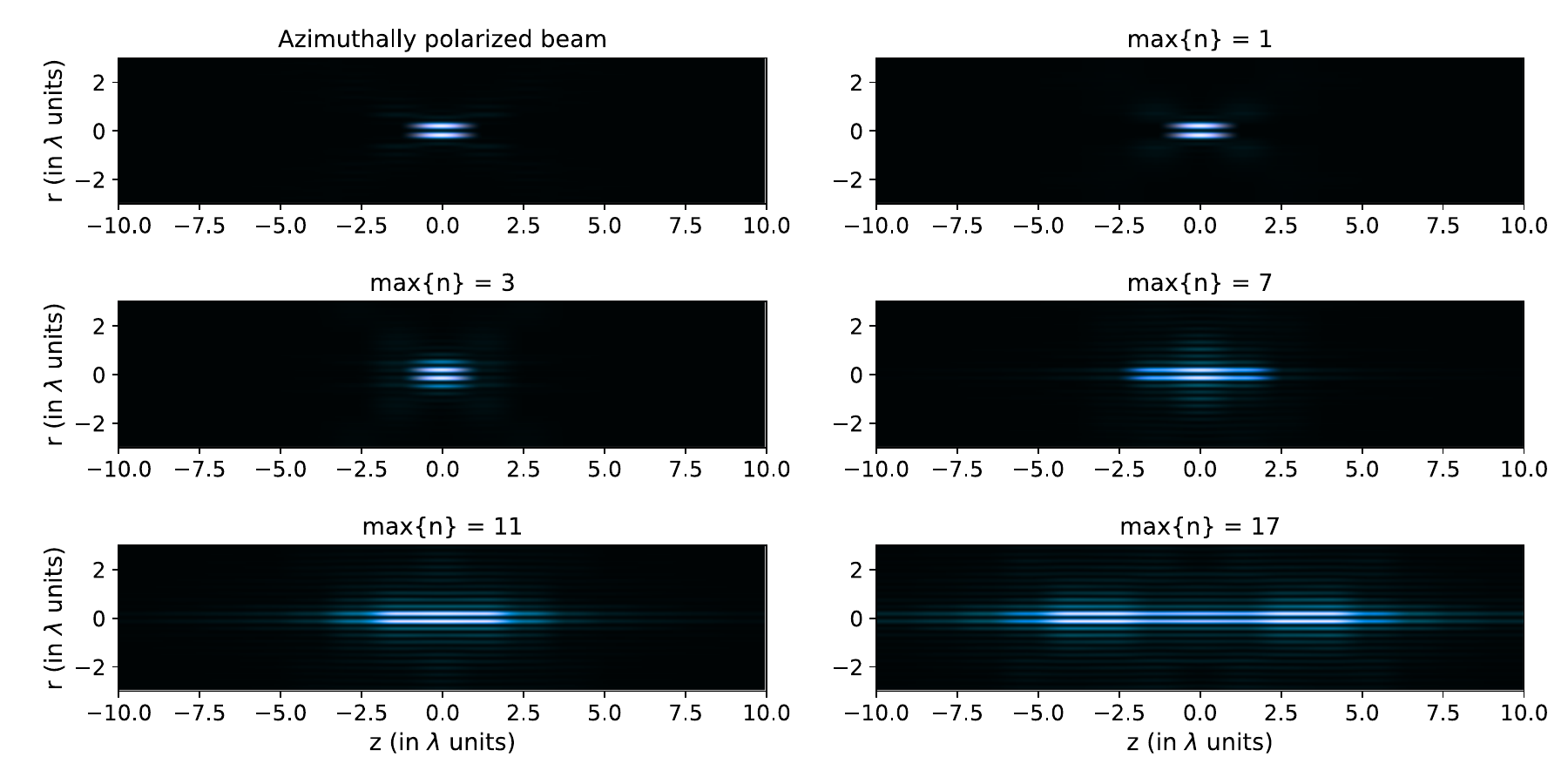}
		\caption{Irradiances $I(r, z)$ for the azimuthally polarized top-hat beam (top left) and the designed beams using $\mathrm{max\{n\}=}$ 1, 3, 7, 11 and 17.}
		\label{fig:propz}
	\end{figure}

	Figure~\ref{fig:propz} displays the propagated irradiance $I(r,z)$ for the top-hat azimuthally polarized beam and the designed beams using $\mathrm{max}\{n\}=1, 3, 7, 11,$ and $17$. Interestingly, the designed beams generate pipe-like structures whose length increases with the number of terms. Taking into account the discussion above, we select $\mathrm{max}\{n\}=7$ and $L = 12$ as beam-design parameters since they represent a reasonable trade-off between depletion resolution, sidelobe production, and longitudinal length. Under these conditions, using DE optimization on Eq.~(\ref{eq:intensitat}), we obtained the set $\alpha_1=0.55, \alpha_3=0.92, \alpha_5=1,$ and $\alpha_7=0.67$. The value $s = 6$ also follows from a compromise between the aforementioned conditions.
    \textcolor{black}{As detailed in Section 1 of Supplementary Material}%  David    R3.2
    , in principle, smaller depletion beams ($s > 6$) could be designed, this would require increasing the number of Zernike terms $n$ to reproduce the target function.
    This, in turn, would necessitate a new assessment of the beam characteristics and a re-optimization of the design parameters.
    These alternative regimes were not explored here, but may offer interesting possibilities, as discussed in the conclusion.
	
	%% ===================================================================
	\section{Estimation of parameter improvements}
	\label{sec:improvements}
	%% ===================================================================
	
	To place the observed reduction of the central minimum on a more quantitative framework, we translated the geometric shrinkage of the doughnuts into expected performance improvements, assuming a STED regime.
	
	To this end, we considered three cases: a standard circularly polarized Laguerre-Gaussian (LG) beam as a reference, the azimuthally polarized beam, and the designed beam. The sharpness of the central intensity null was quantified by approximating the intensity distribution near the optical axis with a parabolic dependence~\cite{Hao:12, Harke:08}, $I(r) \approx \alpha r^2$, where $\alpha$ is the parabolic coefficient describing the curvature of the intensity profile.
	Within this framework, the achievable resolution scaled inversely with the square root of the curvature of the fitted parabola, $d \propto \frac{1}{\sqrt{\alpha}}$.
	Consequently, the relative resolution improvement was obtained as $1 - \sqrt{\alpha_{\mathrm{ref}} / \alpha_{\mathrm{beam}}}$, following the convention of Hao et al.~\cite{Hao:12}, where $\alpha_{\mathrm{ref}}$ denotes the resolution parameter of the reference (worst-performing) beam and $\alpha_{\mathrm{beam}}$ denotes the corresponding parameter of the improved beam.
    
    \definecolor{lightblue}{RGB}{235,245,255}
\begin{table}[h]
	\centering
	\renewcommand{\arraystretch}{1.3}
	\begin{tabular}{lccc}
		\toprule
		\rowcolor{lightblue}
		& {\color[HTML]{2CB92D} \textbf{LG}} & {\color[HTML]{FF03FD} \textbf{Azimuthal}} & {\color[HTML]{3166FF} \textbf{Designed}} \\
		Ring-peak radius $(\lambda)$ & 0.28 & 0.25 & 0.21 \\
		Parabolic coefficient $\alpha$ $(\lambda^{-2})$ & 30.5 & 38.1 & 52.6 \\
		\midrule
		\rowcolor{lightblue}
		\multicolumn{4}{c}{\textbf{Resolution enhancement}} \\
		\quad vs.\ LG  & --- & 10.6\% & 23.9\% \\
		\quad vs.\ Azimuthal & --- & --- & 14.9\% \\
		\midrule
		\rowcolor{lightblue}
		\multicolumn{4}{c}{\textbf{STED power saving}} \\
		\quad vs.\ LG  & --- & 20.0\% & 42.1\% \\
		\quad vs.\ Azimuthal & --- & --- & 28\% \\
		\bottomrule
	\end{tabular}
	\caption{Comparison of simulated STED doughnut beams. The parabolic coefficient
	$\alpha$ is obtained from a quadratic fit to the intensity profile within 30\%
	of each beam's ring-peak radius. Resolution enhancement and STED power saving
	are quoted relative to the standard Laguerre-Gaussian (LG) beam and to the
	azimuthally polarized beam.}
	\label{tab:donut_comparison}
\end{table}

	As noted above, the curvature of the central null also determined how efficiently the depletion threshold was reached. A steeper intensity increase away from the zero implied that the required STED intensity could be achieved with lower overall depletion power.
	% In the high-saturation regime, this led to an approximately quadratic relationship between resolution improvement and power reduction, as reported in previous STED analyses~\cite{Hao:12}.
	\textcolor{black}{Following the analytical approach of Hao et al. \cite{Hao:12}, the depletion intensity distribution is approximated by its quadratic expansion in the vicinity of the central intensity minimum. Under this approximation, the reduction of the central dark region provides an estimate of the corresponding improvement in the theoretical lateral resolution of a depletion microscope.}  %%% Jordi    R2.2

	The resulting figures of merit are summarised in Table~\ref{tab:donut_comparison}. The parabolic approximation was evaluated within 30\% of each beam's ring-peak radius, yielding a coefficient of determination ($R^2$) exceeding 0.99 and a maximum absolute residual below 0.003 (in normalised intensity units) for all three profiles. Transitioning from the circularly polarized LG beam to the azimuthally polarized beam yielded an expected resolution improvement of approximately \textcolor{black}{$10.6\%$} and a corresponding depletion-power reduction of approximately \textcolor{black}{$20\%$}. The engineered beam further enhanced performance, reaching an improvement of approximately \textcolor{black}{$24\%$} in resolution and approximately \textcolor{black}{$42\%$} in depletion power relative to the LG reference. Even the incremental step from the azimuthal to the engineered beam remained significant, corresponding to an additional \textcolor{black}{$15\%$} gain in resolution and a \textcolor{black}{$28\%$} reduction in the required depletion power.
	\textcolor{black}{It should be emphasized that this value corresponds to a theoretical estimate obtained within the standard quadratic approximation of the depletion profile near the focal minimum, as commonly employed in analytical descriptions of STED microscopy, and does not represent an experimentally measured imaging resolution}  %%% Jordi   R2.2
	
	%% ===================================================================
	\section{Beam implementation and experimental results}
	\label{sec:experiment}
	%% ===================================================================
	
\begin{figure}[h!]
		\centering	\includegraphics[width=\linewidth]{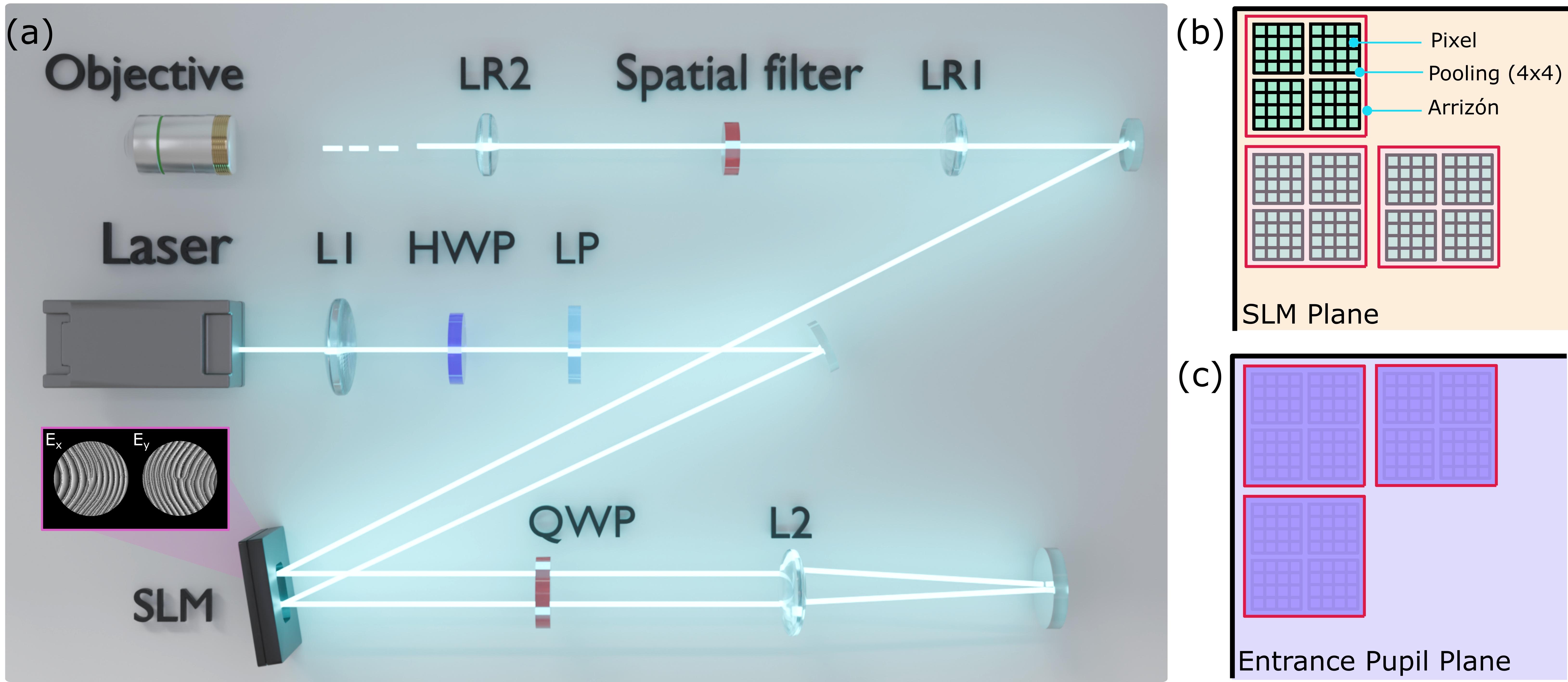}
		\caption{(a) Schematic of the experimental setup. The inset shows the SLM surface divided into two independent regions, each displaying a computer-generated hologram to modulate the two electric field components (rendering generated with Blender, optical package by \textit{Ryo Mizuta Graphics}). (b) Pixel pooling and Arriz\'on encoding at the SLM plane. (c) Smoothing effect of pixel pooling at the entrance pupil of the system.}
		\label{fig:setup_model}
	\end{figure}

	The experimental arrangement used to generate the vector beams is schematically illustrated in Fig.~\ref{fig:setup_model}(a). The setup is based on a reflective, phase-only SLM, which is employed to engineer both the spatial profile and the polarization structure of the beam through computer-generated holograms \cite{maluenda2013reconfigurable}. After modulation, the beam is directed toward a commercial microscope for sample illumination, schematically represented in the figure by an objective lens. Because the SLM can modulate only the electric field component aligned with its modulation axis, the beam-shaping approach relies on a double-pass interaction with the same device \cite{Moreno:12}. This configuration enables sequential modulation of the two orthogonal polarization components, eliminating the need for multiple SLMs or more elaborate optical arrangements.
	
	The input beam is collimated by a first lens (L$_1$), while a half-wave plate (HWP) in combination with a linear polarizer (LP), oriented at $45^\circ$ relative to the modulation axis, ensures that the incident electric field is equally decomposed into two orthogonal polarization components at the SLM surface. During the first interaction with the SLM, only the component of the electric field aligned with the modulation axis acquires the programmed phase profile, while its orthogonal counterpart remains unmodulated. Then, both components are relayed through a 4$f$ system (L$_2$, combined with the mirror). %The subsequent double pass through a quarter-wave plate (QWP) induces a polarization rotation before the beam is re-imaged onto a distinct region of the modulator.
    \textcolor{black}{Since the retardance accumulated over the double pass through the QWP is half a wave, the QWP (fast axis at 45° to the modulation axis) combined with the mirror acts as an effective half-wave plate, interchanging the horizontal and vertical components of the field~\cite{Moreno:12}.}   %%% David    R1.6
    As a result, the polarization component that was previously unaffected becomes aligned with the modulation axis and can be independently phase-shaped. In this way, both orthogonal polarization components are phase-modulated, enabling the generation of the desired vector beam.

	To encode a desired complex (amplitude and phase) optical field using a phase-only SLM, we adopted the double-pixel hologram approach, also known as the Arriz\'on encoding method \cite{arrizon2003complex}. In this scheme, the target amplitude and phase are encoded locally by assigning different phase values to small groups of neighboring pixels on the SLM (Fig.~\ref{fig:setup_model}(b)). Although this pixelated pattern does not directly resemble the target field, its diffraction through a 4$f$ relay system (L$_{R1}$ and L$_{R2}$) causes the contributions from each pixel group to interfere and reconstruct the desired complex field. A spatial filter in the Fourier plane removes high-frequency variations associated with the pixelated structure, leaving only the smooth, low-frequency complex field at the conjugate plane (Fig.~\ref{fig:setup_model}(c)).
	The optical system is configured such that the SLM plane is precisely conjugated to the entrance pupil of the objective, ensuring that the programmed field distribution is accurately relayed to the sample.
	
	Because SLM devices exhibit non-negligible coupling between adjacent pixels (so-called pixel cross-talk), multiple physical pixels are further pooled and driven identically to form a single effective encoding unit~\cite{moreno2021simple}.
    Although this pooling reduces spatial resolution, it significantly stabilizes the local interference process, mitigating cross-talk and improving the fidelity of the reconstructed field.
    \textcolor{black}{We have studied these two competitive effects on increasing the pixel pooling and we found that in the specific conditions of this work (amplitude-only modulation) a $6\times6$ pool size is the best option.
    We detail this study at the Section 2 of the Supplementary Material}    % David     R2.7 
	
	\subsection{Observation of the field at the entrance pupil}
	\label{subsec:pupil}
	
	While the beam has so far been expressed in terms of the radial function $\rho g(\rho)$ (Eq.~(\ref{eq:rhoxg})), its experimental characterization is carried out in terms of light intensity. Therefore, from this point onward, we describe the field through its corresponding intensity distribution, proportional to $\lvert \rho g(\rho) \rvert^2$.
	
\begin{figure}[h!]
		\centering
		\includegraphics[width=\linewidth]{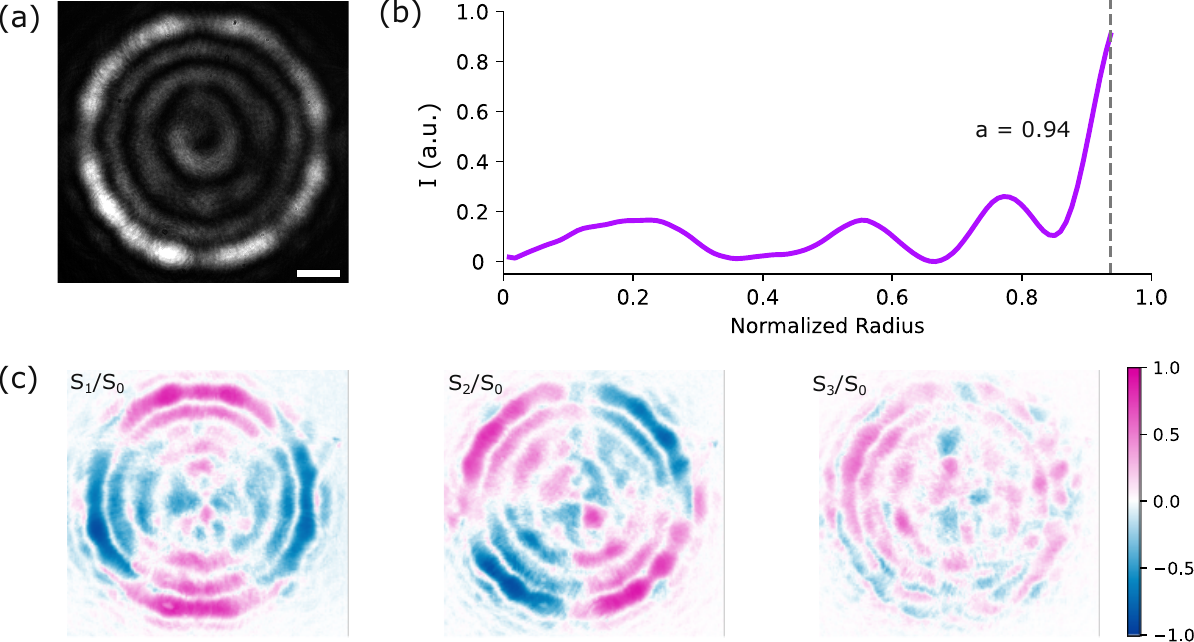}
		\caption{(a) Experimentally measured two-dimensional intensity distributions at the entrance pupil, for the implementation with $n = 7$. Scale bar: \SI{200}{\micro\meter}. (b) Averaged radial intensity profiles extracted from the two-dimensional distribution in (a), plotted as a function of the normalized radius. The vertical gray line and the corresponding label indicates the numerical aperture ($a = 0.94$). (c) Measured normalized Stokes parameter maps $S_1/S_0$, $S_2/S_0$, and $S_3/S_0$ at the entrance pupil for $n = 7$.}
		\label{fig:entrance_pupil}
	\end{figure}

		To verify the correct implementation of the designed pupil pattern, the optical field was observed by removing the objective and imaging the conjugate entrance pupil plane onto a camera. The resulting two-dimensional intensity distribution was measured and directly compared with numerical simulations for the selected implementation ($\textrm{max\{n\}} = 7$), as shown in Fig.~\ref{fig:entrance_pupil}(a).
		
	A direct comparison between experimentally measured and simulated intensity distributions demonstrates good overall agreement. The main features of the radial modulation are well reproduced, confirming the accurate implementation of the designed pupil function. Minor discrepancies are attributed to partial interference between the tails of different diffractive orders, resulting in a slightly reduced intensity uniformity. Fig.~\ref{fig:entrance_pupil}(b) shows the averaged radial intensity profiles, plotted as a function of the normalized radius, confirming that both the number of rings and the radial positions of the intensity zeros are accurately reproduced.
	
	In parallel, polarization-resolved measurements were performed on the same pupil plane using a custom-built polarimeter to characterize the beam's polarization distribution. The resulting normalized Stokes parameter maps % $S_1/S_0$, $S_2/S_0$, and $S_3/S_0$
	are shown in Fig.~\ref{fig:entrance_pupil}(c). The linear polarization components ($S_1$ and $S_2$) exhibit the expected spatial modulation across the pupil, spanning nearly the full range of linear polarization and reaching values close to $\pm 0.9$, with comparable spatial fluctuations (standard deviations of $\pm 0.253$ and $\pm 0.249$, respectively).
	\textcolor{black}{Indeed, the averaged degree of linear polarization is $0.94$}. %%% David   R1.7
	By contrast, the circular polarization component remains weak overall, with a mean value of $S_3/S_0 = 0.04$ and limited spatial variation (standard deviation $\pm 0.126$), although localized regions show deviations up to $\lvert S_3/S_0 \rvert \approx 0.65$. These results confirm the successful generation of the intended azimuthally polarized vector field at the entrance pupil.

	\subsection{Observation of the field at the focal plane}
	\label{subsec:focal}
	
\begin{figure}[h!]
		\centering
		\includegraphics[width=\linewidth]{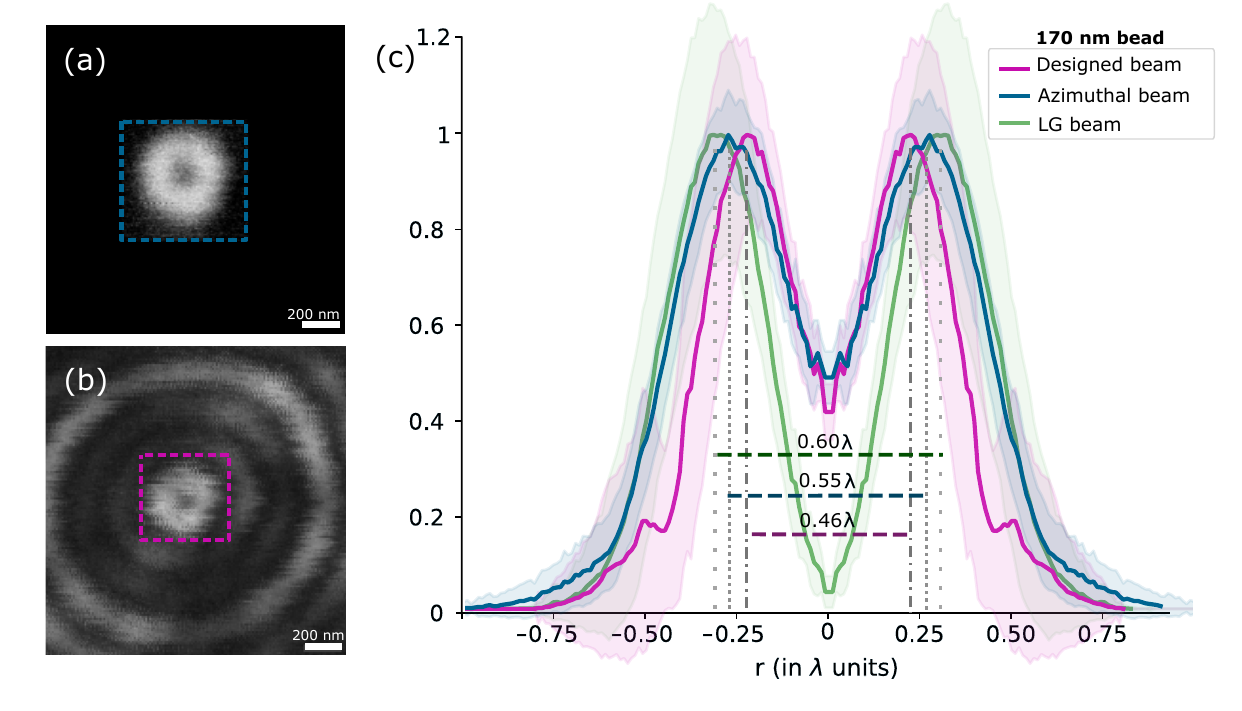}
		\caption{(a) Reconstructed PSF of the nominal azimuthally polarized beam obtained by scanning 170~nm fluorescent beads (5~nm scanning step size). (b) Reconstructed PSF of the engineered beam under the same scanning conditions, showing the central doughnut and surrounding sidelobes. (c) Averaged radial \textcolor{black}{normalized} intensity profiles around the central minimum for the three beams, highlighting the reduced peak-to-peak separation of the designed beam\textcolor{black}{; shaded bands are ±1 standard deviation of the intensity across the azimuthal angle at each radius (angular uniformity, not a formal error on the ring positions).}}  %%%  David / COstanza    R2.3 / R4.4
		\label{fig:170}
	\end{figure}

	As shown in Section~\ref{sec:design}, the tailored pupil-plane amplitude and azimuthal polarization distributions are expected, under high-NA focusing, to generate a doughnut-shaped focal field governed by Bessel-function weighting.
    To experimentally characterize this focal distribution, sub-diffraction fluorescent beads are raster-scanned across the focal plane while the emitted fluorescence is recorded at each position.
    Because the bead size is significantly smaller than the diffraction-limited spot, the detected fluorescence at each scan position provides a good approximation of the local excitation intensity.
    By integrating the recorded signal over the two-dimensional scan grid, an experimental map of the effective point spread function (PSF) is obtained~\cite{Cole2011MeasuringAI}.
    This approach provides a direct experimental representation of the focal field associated with each beam configuration and allows quantitative comparison between different designs (Fig.~\ref{fig:170}).
	
	To probe the focal field, we first scanned fluorescent beads with a diameter of 170~nm.
    These beads provide strong emission and high photostability, ensuring that the focal intensity distribution can be reliably captured with short exposure times and small scanning steps.
	The reconstructed PSFs obtained from the bead scans are shown in Fig.~\ref{fig:170}(a) and (b) for the nominal azimuthally polarized beam and the designed beam, respectively.
    Both beams exhibit the expected doughnut-shaped focal structure.
    In addition to the central minimum and main intensity ring, the designed beam (Fig.~\ref{fig:170}(b)) clearly reveals the presence of surrounding sidelobes, consistent with the predicted Bessel-like \textcolor{black}{beams~\cite{durnin1987diffraction}}.
	To quantify the central narrowing, the averaged radial \textcolor{black}{normalized}    % David     R2.3
    intensity profiles around the central minimum are extracted and plotted in Fig.~\ref{fig:170}(c), where a circularly polarized Laguerre-Gaussian (LG) beam was also included as a comparison.
    The profiles, obtained from scanning with a 5~nm step size, reveal that the designed beam produces a tighter central doughnut compared with the nominal azimuthally polarized beam.
    Specifically, the experimental peak-to-peak distance is reduced from $0.55\,\lambda$ for the reference beam to $0.46\,\lambda$ for the implemented beam, corresponding to a relative improvement of approximately 16\%.
    For comparison, the circularly polarized Laguerre-Gaussian reference exhibits a peak-to-peak distance of approximately $0.60\,\lambda$, reduced by approximately 8\% for the azimuthally polarized beam, while the engineered beam achieves a reduction of up to 23\% relative to this LG configuration.
	 The optical power was not matched between the cases due to the reduced efficiency of the engineered beam, as discussed below.
    Nevertheless, the measured reduction in peak-to-peak distance reflects an intrinsic geometric property of the beam: for a given peak intensity of the depletion ring, a narrower central zero would translate into enhanced lateral resolution.
	
	The 170~nm beads used in the initial measurements provided a reliable characterization of the focal intensity distribution and enabled comparison with reference beams. However, their size is comparable to the expected width of the central minimum and can partially obscure its true suppression. This effect is visible in Fig.~\ref{fig:170}(c), where the wider LG doughnut exhibits a more clearly defined central zero due to its larger diameter.
	\textcolor{black}{The use of 170~nm beads was motivated}   %%% David   R1.8
    by the reduced focal intensity of the designed beam, which results from its inherent structure and from cumulative design choices at the pupil plane. Specifically: the Bessel-like nature of the beam, with its characteristic sidelobes, tends to distribute energy across the pupil. In addition, phase-only holographic encoding based on the Arriz\'on method further redistributes the optical power into undesired diffraction orders. Finally, a top-hat amplitude profile was applied at the entrance pupil to compensate intensity modulation from the Gaussian illumination, producing a uniform intensity distribution at the cost of reduced transmitted power. Collectively, these factors lower the fluorescence signal at the focus while optimizing the spatial and polarization structure of the beam. The 170~nm beads therefore ensured sufficient signal for robust initial measurements.
	
\begin{figure}[th!]
		\centering
		\includegraphics[width=\linewidth]{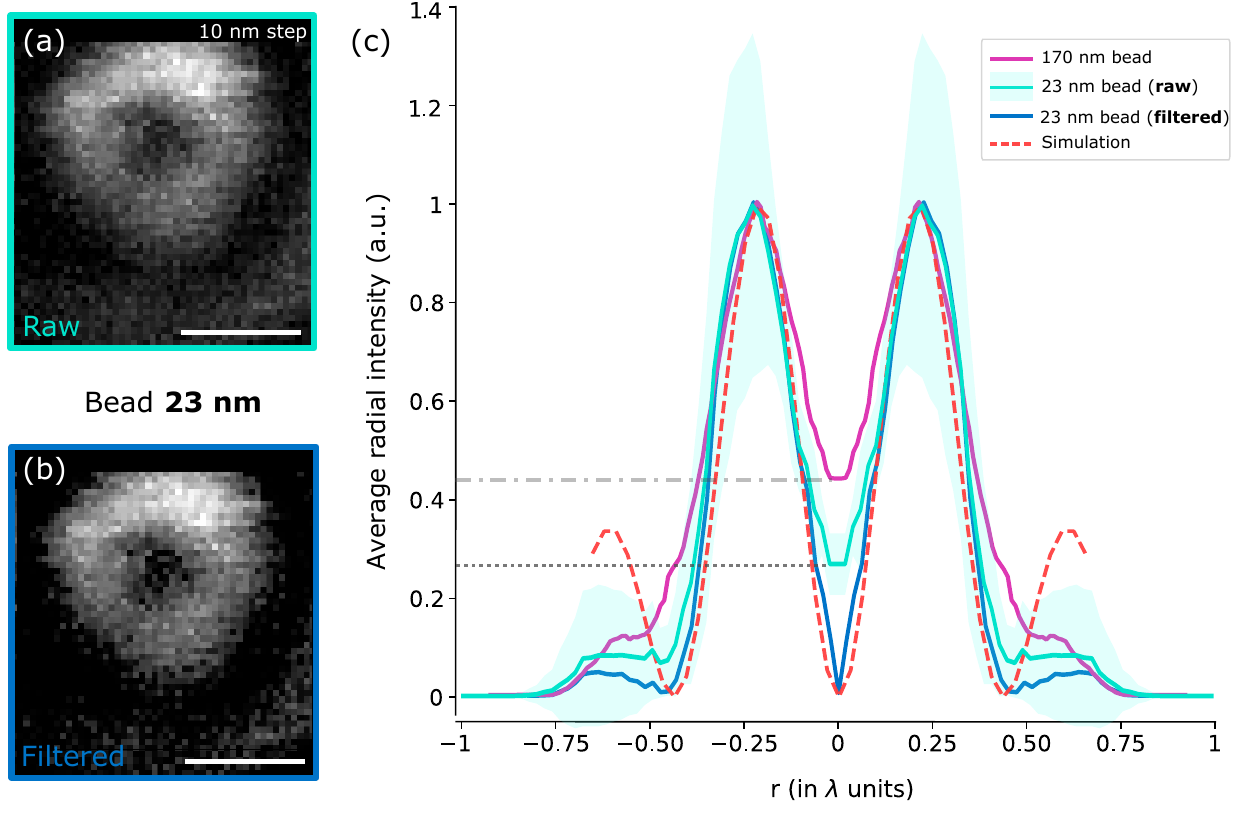}  %%% R4.4
		\caption{Characterization of the central optical zero using 23~nm fluorescent beads.
			(a) Reconstructed PSF of the designed beam obtained from a 23~nm bead scan with 10~nm step size (raw data).
			(b) Same reconstruction after bilateral filtering. Scale bar: 200~nm.
			(c) Averaged radial intensity profiles of the raw and filtered 23~nm scans, compared with the profile obtained using 170~nm beads (magenta) and with the simulated beam (dotted red line), highlighting the improved spatial confinement of the central zero\textcolor{black}{; shaded bands are ±1 standard deviation of the intensity across the azimuthal angle at each radius (angular uniformity, not a formal error on the ring positions).}}  %%% Costanza    R4.4
		\label{fig:small_beads}
	\end{figure}

	To quantitatively assess the on-axis intensity minimum, we repeated the focal scans using sub-diffraction fluorescent beads of 23~nm, well below the expected size of the central zero. Figure~\ref{fig:small_beads}(a) shows the reconstructed PSF obtained from the 23~nm bead scan (10~nm step size), displaying the central doughnut of the designed beam.
	The corresponding averaged radial intensity profile is plotted (cyan curve) in Fig.~\ref{fig:small_beads}(c). For comparison, the profile obtained with 170~nm beads (shown in magenta) and the simulated beam profile (dotted red line) are also included. The 23~nm measurement reveals a more sharply defined on-axis minimum, with a narrower profile and steeper intensity walls compared with the 170~nm case, confirming that the larger beads partially fill the central minimum and underestimate its true suppression.
	
	Scanning with 23~nm beads remains experimentally demanding, particularly given the relatively low efficiency of our designed beam. Additionally, the smaller beads are highly light-sensitive, which constrains both the scanning area and the acquisition speed. The increased step size was therefore chosen to mitigate photobleaching while maintaining adequate spatial sampling. Longer exposure times, while necessary to collect sufficient signal, also increase the contribution of background noise, making it more difficult to directly observe the true depth of the central zero.
    
	To explore the potential for a cleaner measurement, we applied post-processing techniques aimed at reducing high-frequency noise while preserving the key features of the beam.
	\textcolor{black}{Among the methods tested, we focus here on the bilateral filter, as implemented in the scikit-image library~\cite{vanderWalt2014ScikitImage}. This edge-preserving filter averages pixels according to both their spatial proximity ($\sigma_{\mathrm{spatial}}$) and their intensity similarity ($\sigma_{\mathrm{color}}$). We used $\sigma_{\mathrm{color}} = 0.05$ and $\sigma_{\mathrm{spatial}} = 40$~px on the $10$~nm/pixel grid. The value of $\sigma_{\mathrm{color}}$ was chosen to match the measured noise amplitude ($\sigma \approx 0.05$ on the normalized intensity scale, corresponding to a signal-to-noise ratio of $\mathrm{SNR} \approx 20$). Since this value is well below the ring-to-null intensity contrast, the filter suppresses fluctuations at the noise level while preserving the much larger intensity transition between the ring and the central minimum. The choice of $\sigma_{\mathrm{spatial}}$ reflects the sparse sampling of the scan.
	The filtered reconstruction is shown in Fig.~\ref{fig:small_beads}(b), and the corresponding radial profile (blue curve, Fig.~\ref{fig:small_beads}(c)) provides a denoised representation of the raw data (magenta curve) that is in closer visual agreement with the simulation. To verify that the filtering does not modify the beam profile, we evaluated a wide range of filter parameters. Across this range, both the ring position and the ring-to-ring (peak-to-peak) width remain constant within the measurement sampling, confirming that the observed narrowing is intrinsic to the beam rather than an artifact of the filtering.
	In contrast, the apparent depth of the central minimum shows a residual dependence on the filter parameters, as well as on the normalization and centering procedures. For this reason, the filtered profile is used only as a denoised representation of the measurement, and no quantitative value of the  absolute suppression is extracted from it. Instead, the conclusion that the true central minimum is more strongly suppressed than suggested by the $170$~nm measurements is based on the smaller probe size.}  %%% Costanza    R1.2

	%Among the methods tested, we focus here on the bilateral filter, as implemented in the scikit-image image processing library \cite{vanderWalt2014ScikitImage}. The bilateral filter is an edge-preserving smoothing algorithm that performs weighted averaging based on both spatial proximity and intensity similarity, controlled respectively by two sigma parameters. In our case, we used a very large spatial-sigma, effectively averaging over nearly the entire image, while a small intensity-sigma constrained the averaging to pixels with similar intensity values, thereby enhancing contrast at intensity transitions. The filtered reconstruction is shown in Fig.~\ref{fig:small_beads}(b).
	%The averaged radial intensity profile after filtering (blue curve in Fig.~\ref{fig:small_beads}(c)) shows a sharper and more spatially confined central zero, bringing it into closer agreement with the simulated profile. This indicates that noise contribution in the raw measurement partially obscures the true level of destructive interference at the beam center.
	%Overall, these results confirm that the designed beam supports a more spatially confined optical zero than initially inferred from measurements performed with larger beads.

	\subsection{\textcolor{black}{Comparison between simulated and experimental beams}}
    \label{subsec:exp_vs_sim}

\begin{figure}[t]
        \centering
        \includegraphics[width=\linewidth]{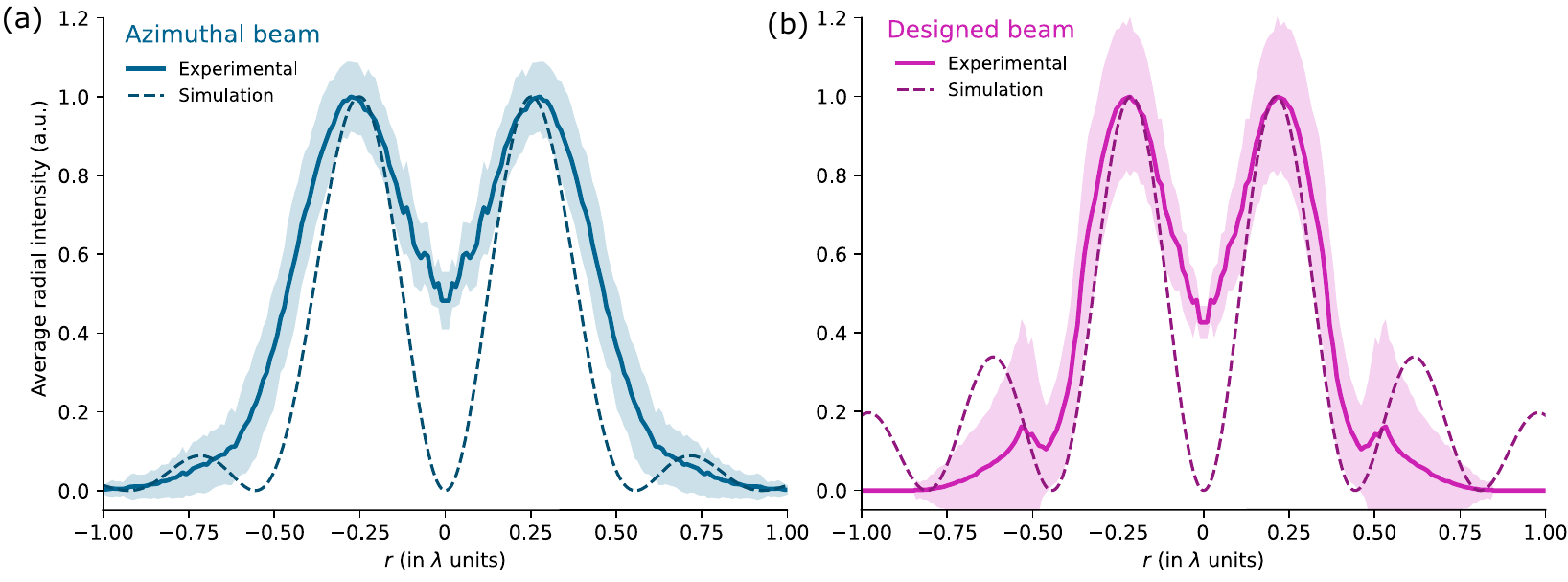}
        \caption{Comparison between the simulated (dotted lines) and experimentally
            reconstructed (solid lines) averaged radial intensity profiles for (a) the azimuthally polarized and (b) the designed beams. The shaded bands indicate
            $\pm1$ standard deviation of the intensity over the azimuthal angle at each
            radius.}
        \label{fig:exp_vs_sim}
    \end{figure}

    \textcolor{black}{To assess how faithfully the engineered field is reproduced in practice, we compared the experimentally (using beads of 170~nm) reconstructed radial intensity profiles with the corresponding vectorial simulations for both, the designed and the azimuthally polarized beams (Fig.~\ref{fig:exp_vs_sim}).
    %The experimentally reconstructed radial intensity profiles, obtained from the 170 nm bead scans, are shown together with shaded bands denoting $\pm 1$ standard deviation of the intensity across the azimuthal angle at each radius, reflecting the angular uniformity of the ring.
    }

    \textcolor{black}{For the designed beam, the measured profile closely follows the prediction: the main intensity ring is located at $0.215\,\lambda$ experimentally, compared with $0.211\,\lambda$ in the simulation, an agreement to within $\sim\!2\%$.
    The azimuthally polarized reference beam, generated with the same platform, likewise reproduces the expected doughnut, with its ring at $0.277\,\lambda$ experimentally versus $0.252\,\lambda$ in simulation. This good correspondence serves as an independent validation that the optical system and the reconstruction procedure behave as intended.}    %%% Costanza     R4.2

    \textcolor{black}{Taken together, these results confirm that the beam reshaping predicted by the vectorial model is faithfully reproduced in the experiment. Beyond the sub-2\% agreement in the absolute ring positions reported above, the relative figures of merit are also recovered: the engineered beam yields a lateral resolution improvement of approximately 16\% with respect to the azimuthally polarized reference, in close agreement with the 14.9\% resolution enhancement predicted analytically for the same pair of beams (Table~\ref{tab:donut_comparison}).}     %%% David R4.2   per tancar la secció i la part de RESULTS del paper (prior to discussion)

    %This consistency between the measured geometric narrowing and the model-based estimate indicates that the anticipated gain in lateral resolution — and, through the quadratic relation near the central minimum, the associated reduction in the required depletion power — is indeed attainable with the implemented field. A direct experimental quantification of the depletion-power saving would nonetheless require operation in a full STED or RESOLFT configuration, which is left for future work.    %%% Això si de cas a discussion

	%% ===================================================================
	\section{Discussion}
	\label{sec:discussion}
	%% ===================================================================
	
	The results presented in this work demonstrate that super-oscillatory pupil engineering, combined with full vectorial polarization control, is a viable strategy for tightening the central zero of azimuthally polarized depletion beams. While super-oscillatory designs are generally considered disadvantageous for conventional imaging due to the redistribution of energy into sidelobes, depletion microscopy operates under a different figure of merit: while resolution depends on several factors including depletion power and aberrations \cite{Galiani:12}, the spatial confinement and steepness of the central intensity minimum constitute a fundamental and engineerable parameter, motivating the present approach. At the same time, sidelobe energy is not without consequence, as it reduces the optical power available for fluorescence depletion and therefore affects the overall efficiency of the process. The beam presented here was designed with these competing requirements in mind, seeking to enhance central confinement while keeping sidelobes at a manageable level. By approximating the intensity near the optical axis with a parabolic dependence \cite{Hao:12}, the curvature of the central minimum could be directly linked to both achievable resolution and depletion efficiency.
    Within this framework, the analytical engineered beam is predicted to outperform both the standard circularly polarized Laguerre-Gaussian and azimuthally polarized beams, yielding a resolution enhancement of \textcolor{black}{$\sim 24\%$} and a depletion power reduction of \textcolor{black}{$\sim 42\%$} relative to the LG reference, along with an additional \textcolor{black}{$\sim15\%$} resolution gain over the azimuthally polarized beam.
	
	Within the vectorial Richards-Wolf framework, the pupil function was expressed as a truncated Zernike expansion and the coefficients were optimized to approximate a scaled target distribution $|\mathrm{J}_1(snr)|^2$. The optimization was performed over a design window of $L = 12\lambda$, which was found to minimize sidelobe production compared to smaller domains, and the expansion was limited to a maximum radial order of $n = 7$. This combination represents the best compromise between central confinement and sidelobe suppression: increasing $n$ beyond 7 improves the approximation of the target function but progressively strengthens sidelobes and extends the axial beam structure. The scaling factor was set to $s = 6$, selected as the smallest value that yields a measurable improvement in central confinement. In principle, larger values of $s$ would produce progressively narrower depletion regions, but at the cost of a longer axial beam extent, eventual stronger sidelobes in the optimization window, while also introducing significant challenges for experimental characterization and control.
	
	Experimentally, the vector beams were generated and characterized using a phase-only SLM in a double-pass configuration, which provided independent control of the two orthogonal polarization components and accurate relay of the field to the sample plane. The correct implementation of the designed pupil function was first verified at the entrance pupil plane, where the measured radial intensity profiles, including the number of rings and the positions of the intensity zeros, were found to be in good agreement with simulations.
	Polarimetric measurements of the normalized Stokes parameters further confirmed the successful generation of an azimuthally polarized field.
	
	A dedicated optimization of the holographic encoding was performed to maximize the fidelity of the central optical zero at the focus. Because pixel cross-talk in reflective SLMs distorts fine spatial features and degrades the destructive interference required for a narrow, well-defined on-axis null, physical pixels were pooled into larger effective encoding units. Testing pooling factors from $2 \times 2$ to $8 \times 8$ revealed that a $6 \times 6$ configuration provides the optimal balance: it sufficiently suppresses cross-talk to deepen the central zero while retaining enough spatial resolution to accurately reproduce the designed pupil pattern. Beyond this threshold, undersampling begins to degrade the reconstructed field.
	
	Focal scans with 170~nm fluorescent beads confirmed the predicted beam reshaping: the doughnut diameter was reduced from $0.55\lambda$ for the conventional azimuthally polarized beam to $0.46\lambda$ for the engineered beam, corresponding to a resolution improvement of approximately $16\%$.
	%Although the optical power was not matched between the two cases due to the reduced efficiency of the engineered beam, this reduction in peak-to-peak distance reflects an intrinsic geometric property of the focal field.   %%% David: Trec això perque ara ja coincideix amb el numeric
    For a given depletion-ring intensity, such narrowing would directly translate into improved lateral resolution.
	The 170~nm beads were selected for this initial characterization because the reduced focal intensity of the designed beam required probes with strong emission and high photostability to ensure robust signal acquisition. However, since the bead diameter is comparable to the expected width of the central minimum, these measurements inevitably lead to partial filling of the zero and an underestimate of its true suppression. Complementary measurements with 23~nm beads, whose smaller size minimizes this probe-induced filling, revealed a more sharply defined on-axis minimum, with a narrower profile and steeper intensity walls, in closer agreement with the simulated profile.
	\textcolor{black}{Edge-preserving bilateral filtering was applied to the $23$~nm scans to reduce background noise and improve visualization.  We verified that the filtering preserves the ring geometry, and therefore the measured doughnut width, over a broad range of filter parameters. Accordingly, the filtered images are used only for visualization, while all quantitative analysis is based on the unfiltered data.}  %%% Costanza     R1.2
	%Edge-preserving bilateral filtering of the 23~nm scans further confirmed the robustness of the optical zero by reducing background noise contributions while preserving the key intensity features of the beam.
	
	%% ===================================================================
	\section{Conclusion}
	\label{sec:conclusion}
	%% ===================================================================
	
	%This work demonstrates that super-oscillatory pupil engineering, implemented on a reconfigurable SLM platform, provides an experimentally viable route to sub-diffraction reshaping of azimuthally polarized depletion beams.
	\textcolor{black}{This work demonstrates that super-pupil engineering can further improve the focal confinement of conventional azimuthally polarized depletion beams.
    By tailoring the radial amplitude distribution at the entrance pupil to approximate a Bessel-type target function, we obtain a focal field with a tighter central doughnut than that of conventional circularly or azimuthally polarized beams.
}  %%% Jordi   R1.1       % reworded from:   By tailoring the radial amplitude distribution at the entrance pupil, we experimentally obtain a more tightly confined central intensity minimum while preserving the vectorial focusing properties that make azimuthally polarized beams attractive for fluorescence depletion microscopy.
	Specifically, it has been shown:
	(i) that super-oscillatory pupil engineering can be used to achieve sub-diffraction reshaping of azimuthally polarized doughnut beams suited for depletion-based super-resolution microscopy;
	(ii) that a reconfigurable, phase-only SLM platform, based on a double-pass configuration, can reliably reproduce not only canonical depletion beams but also more complex engineered vector fields; and
	(iii) that careful characterization and optimization of hologram encoding enable controlled enhancement of local intensity zeros in experimentally generated beams.

	\textcolor{black}{The proposed platform also demonstrates that complex vector depletion beams can be generated reproducibly using phase-only spatial light modulation. Careful optimization of the hologram encoding, including controlled pixel pooling, proved essential to compensate for SLM cross-talk and to accurately reproduce the designed pupil function, enabling the experimental realization of superpupil beam designs.}  %%% Jordi   R2.7

    \textcolor{black}{It should be noted that the double-pass SLM configuration employed in this work is a convenient and versatile platform for generating arbitrary vector beams, but it is not a requirement for implementing the proposed design. In practice, the beam could be realized using fixed optical elements, for example by combining an azimuthal polarization converter with a pupil transmission mask.}  %%% Mario    R2.1

	Several directions remain open for future investigation. To overcome the throughput limitation identified above, future work will explore iterative phase-retrieval approaches such as the Gerchberg-Saxton algorithm \cite{gerchberg72practical}, which generates full-phase holograms capable of directing a significantly larger fraction of the incident light into the target field.
	From a computational standpoint, a more systematic exploration of the design space, spanning a wider range of scaling factors $s$ and maximum Zernike orders $n$, as well as different design window sizes $L$, would provide a comprehensive characterization of the trade-offs between central confinement, sidelobe intensity, and total optical power at the focus.
	From an application standpoint, an important next step is the implementation of the engineered beams in experimental RESOLFT or STED imaging to quantify the achievable resolution enhancement and depletion power reduction in biological samples, thereby enabling validation of the analytical predictions presented here.
	%Such measurements would provide a direct assessment of the benefit conferred by the tighter central zero compared to conventional circularly and azimuthally polarized beams.
	\textcolor{black}{These studies will enable experimental validation of the analytical predictions presented here in complete RESOLFT imaging experiments and will allow us to assess the robustness of the proposed beam under realistic optical aberrations and imaging conditions.}  %%% Jordi     R2.8

	In addition, the axial structure of the engineered beams would deserve further investigation. The pipe-like elongated intensity distributions produced by higher-order designs could be exploited for volumetric or multi-plane depletion schemes, although their impact on depth of field and three-dimensional resolution remains to be characterized.
	\textcolor{black}{Also, a systematic study of the robustness of the proposed beam under controlled optical aberrations will be important to assess its practical performance in realistic microscopy systems.}  %%% Jordi   R2.8
	
	%Overall, this work establishes a direct link between theoretical beam design, experimental realization, and quantitative validation. It explores an alternative route to established depletion beam generation using a programmable platform, with the final goal of preserving key beam characteristics, such as azimuthal polarization, while enabling new strategies, such as sub-diffraction central pipes from super-pupil designs.   R1.1
	\textcolor{black}{Overall, this work establishes a direct link between vectorial beam engineering, super-pupil design, and experimental implementation on a programmable optical platform. Rather than introducing an alternative depletion beam concept, the proposed approach builds upon the favorable focusing properties of azimuthally polarized beams and demonstrates that their central intensity minimum can be further confined through super-pupil engineering. We believe that this beam-engineering strategy provides a flexible framework for the future development of optimized depletion beams for STED and RESOLFT microscopy.}  %%% Jordi   R1.1

	%% ===================================================================
	\section*{Funding}
	%% ===================================================================
	This work was supported by Grant PID2022-136796OB-I00 funded by Ministerio de Ciencia e Innovaci\'on, Spain, MCIN/AEI/10.13039/501100011033.
	
	\section*{CRediT authorship contribution statement}
	Costanza Agazzi: Investigation, Formal analysis, Data curation, Writing - original draft, Visualization. 
	Nick Toledo-García: Software, Resources, Investigation, Writing - review \& editing.
	Estela Martín-Badosa: Conceptualization, Writing - review \& editing, Supervision, Project administration, Funding acquisition.
	Mario Montes-Usategui: Conceptualization, Methodology, Resources, Funding acquisition.
	Jordi Tiana-Alsina: Conceptualization, Methodology, Writing - original draft, Supervision. 
	David Maluenda: Conceptualization, Writing - review \& editing, Supervision. 
	Rosario Martínez-Herrero: Conceptualization, Writing - review \& editing. 
	Artur Carnicer: Conceptualization, Software, Writing - original draft, Supervision.
	
	\section*{Declaration of competing interest}
	The authors declare no conflicts of interest.
	
	\section*{Data availability}
	The code for generating Figs.~2--5 and Figs.~S1--S2 can be found at \url{https://github.com/WavefrontEngUB/VeReady}.
	
	%% Bibliography
	\bibliographystyle{elsarticle-num}
	\bibliography{refs}
	
\end{document}